# Comparison of Check-All-That-Apply and Adapted-Pivot-Test methods for wine descriptive analyses with a panel of untrained students

Sylvain Nougarede[1,3], Alice Diot[1], Elie Maza[1], Alain Samson[2], Valérie Olivier-Salvagnac[3], Soline Caillé[4], Olivier Geffroy[5], Christian Chervin[1].

[1] ENSAT, Toulouse INP, LRSV, GBF, CNRS, UPS, Université de Toulouse, Auzeville-Tolosane, France

[2] UE Pech Rouge, INRAE, Gruissan, France

[3] UMR AGIR, Toulouse-INP, INRAE, Auzeville-Tolosane, France

[4] UMR SPO, Univ Montpellier, INRAE, Institut Agro, Montpellier, France

[5] PPGV, INP-PURPAN, Université de Toulouse, Toulouse, France

**Abstract**

The Check-All-That-Apply (CATA) method was compared to the Adapted-Pivot-Test (APT) method, a recently published method based on pair comparisons between a coded wine and a reference sample, called pivot, and using a set list of attributes as in CATA. Both methods were compared using identical wines, correspondence analyses and Chi-square test of independence, and very similar questionnaires. The list of attributes used for describing the wines was established in a prior analysis by a subset of the panel. The results showed that CATA was more robust and more descriptive than the APT with 50 to 60 panelists. The p-value of the Chi-square test of independence between wines and descriptors dropped below 0.05 around 50 panelists with the CATA method, when it never dropped below 0.8 with the APT. The discussion highlights differences in settings and logistics which render the CATA more robust and easier to run. One of the objectives was also to propose an easy set-up for university and food industry laboratories.

**Practical applications**

Our results describe a practical way of teaching and performing the CATA method with university students and online tools, as well as in extension courses. It should have applications with consumer studies for the characterization of various food products. Additionally, we provide an improved R script for correspondence analyses used in descriptive analyses and a Chi-square test to estimate the number of panelists leading to robust results. Finally, we give a set of data that could be useful for sensory and statistics teaching.

## 1. INTRODUCTION

Descriptive analyses of food by consumers are always challenging (Ares and Jaeger, 2015), due to high variability, but are necessary, for example, to correlate the hedonic scores with words commonly used by consumers to describe their sensory perception, or to correlate repurchase intentions with the sensory perception.

We recently published in this journal the development of a descriptive analysis adapted to wine consumers (Beaulieu et al. 2022), we called it Adapted Pivot Test, as it is based on a Pivot© Profile method initially developed with wine experts (Thuillier et al. 2015). These methods belong to the reference-based methods (Valentin et al. 2012) and present the advantage to run pair comparisons between a coded wine and a reference wine, called 'pivot', made of a combination of equal proportions of the tested wines. We found this concept interesting as it could limit the variability of consumer judgements, giving them a reference. And indeed, the APT provided an interesting set of correspondence analyses allowing the characterization of white and red wines by young untrained students (Beaulieu et al., 2022), who represent a subset of young consumers. Researchers recently published the development of a similar method, called "PP+CATA" or "Pivot-CATA", which they tested with whey-based fermented beverages (Miraballes et al., 2018) and instant black coffee (Wang et al. 2023).

The APT method led to interesting results and notably a product map in accordance with the expected sensory characteristics of the wines. However, the validation of this new method would require comparing it with other alternative methods such as CATA that can also be used with consumers. This comparison has been the objective of the experiments described in this article.

CATA analyses were developed in the middle of the 2000's, and have been extensively used in sensory analyses in the past decade, with more than 400 articles found in Web of Science when crossing 'CATA' and 'sensory' keywords in a Topic search. Gaston Ares and Sarah Jaeger, among other authors, have published series of studies, describing the method (Ares and Jaeger, 2015), testing the optimal number of terms to list in a CATA questionnaire (Jaeger et al., 2015), and comparing the CATA analysis to a classical descriptive analysis using intensity scales (Jaeger et al, 2020), among other works.

Over the last decade, wine quality assessments using CATA analyses were reported from all over the word. Vidal et al. (2018) used CATA to assess red wine astringency in Uruguay; Rinaldi et al. (2021) performed similar studies in Italy with different cultivars; Phan et al. (2022) used CATA to check the effects of various processes on red wine quality in the USA; Yang and Lee (2021) used CATA with Korean consumers and red wine from various origins; Lezeata et al. (2017) used CATA in a partner study between Chilean and Norwegian teams about Sauvignon blanc; Brand et al. (2020) used CATA with South African white wines; and Ruppert et al. (2021) used it with apple wines in Austria. The list is obviously non exhaustive.

Thus, we will compare here the CATA and APT methods using the same correspondence analysis method, the same wines and similar questionnaires and panelists, then comment on robustness, logistics and other aspects.

## 2. MATERIAL AND METHODS

### 2.1. Wines

White wines were made in 2021 in the South of France, Domaine de Pech Rouge, INRAE, Gruissan, South of France. All experiments were performed with a total of 8 bottles of 750 mL for each wine. The wine names used in this study are: *VIOGNIER1* and *VIOGNIER2* for an identical wine, made with this *V. vinifera* cultivar, but presented twice, under different codes, to check if it would be analyzed as an identical wine; *MUSCAT* for a wine made from an interspecific hybrid, for which the last cross was *V. vinifera* cv. Muscat de Hambourg (INRAE label: G5 [3197-81]); *ITALIA* for a wine made from an interspecific hybrid, for which the last cross was *V. vinifera* cv. Italia (INRAE label: G9 [3196-57]); *CHAS-SPUR* and *CHAS-MP* for wines made from an interspecific hybrid, for which the last cross was *V. vinifera* cv. Chasan (INRAE label: G9 [3196-57]); *CHAS-SPUR* means spur pruning with limited fruit load; *CHAS-MP* means minimal pruning with high fruit load. The physicochemical characteristics of the five wines are given in Supplemental Table 1.

### 2.2. Sensory panel

The three sessions were run with very similar panels, composed by students of the ENSAT Agronomy School, average age 21.6 years, with a total panel of 42 women and 25 men. The details of the CATA and APT sensory panels regarding number of panelists, age and sex are given in Supplemental Table 2. Both methods were tested in two separate sessions, on two different days. A first session to generate 'attributes' ('descriptors' being restricted to trained panels) was performed with 30 panelists, randomly chosen from the total panel. All panelists were enrolled in the sensory sessions, based on the willingness to participate, and without particular training in wine sensory analysis. At the beginning of the sessions, all participants gave informed and written consent.

### 2.3. Sensory methods

In a preliminary session, the five wines were served to each panelist in random order, coded with 3-digit numbers, and the panelists were asked to write up to four attributes (preferably adjectives), describing their olfactory and gustatory perception. The words were then lemmatized and a triangulation was performed by three oenologists in order to group the words into main families, as previously described (Symoneaux et al., 2012), then discrepancies were discussed until consensus was obtained. For the forthcoming CATA and APT tests, it was chosen to work with the 17 words representing the highest sums of citation occurrence. The details are given in Supplemental Table 3. We chose a maximum of 17 words, as a number of 10 to 17 words seems optimal in CATA tests to avoid a ''dilution'' effect of the responses (Jaeger et al., 2015).

The CATA set-up was derived from the method published by Ares and Jaeger (2015), with the following adaptations to run it with a university class. The paper questionnaire was presenting the consent form on one side, and the instructions plus the table to be filled on the other side. This latter is shown in Supplemental Table 4. The attributes were presented in random orders from top to bottom. To be practical, only 10 different orders were generated for attributes, and were distributed equally among panelists, see Supplemental Table 3. It was also found practical to write a discrete indication of this order on the questionnaire (e.g. (o3) for the third random order), for a faster typing-in in the corresponding tab of the online shared table, when treating the data. Each panelist had one glass and received 20 ml of a coded wine at a time, wine at 15°C, in transparent glasses. The wines were served using random orders for each panelist. Wine codes were random 3-digit numbers. Each wine was served with a small graduated vial, brought by the server from the preparation area, allowing to limit the number of glasses per session, limiting also the risk of breaking a glass, or a bottle while circulating in the tasting room, and also limiting finger contacts on the glasses. For practical reasons, a specified digit of 3-digit codes was indicative of the wine product which was served. Only the persons in charge of serving were aware of this code correspondence. This allows the students to serve the wines without a correspondence sheet, which often creates an additional source of error. Example: if the specified digit is the second of a 3-digit code, the wine number 4 was served with various codes such as: 145, 348, 043, 947, etc… For repetitive sessions with the same panel, 3- to 6-digit numbers can be used, with specific digits at different places, which strongly limit the ability of panelists to find the coding system. Then the panelist smelled and tasted the wine before checking the attributes corresponding to their perception. At the end, they were asked to give a hedonic score from 1 (I do not like it) to 9 (I like it very much), as proposed in combination to some CATA trials (Ares and Jaeger, 2015). Between each wine, they had to rinse their glass and their mouth with water.

The APT set-up was derived from the method published by Beaulieu et al. (2022), with the following adaptations. Each panelist had two glasses and received two wines at a time, 20 mL of each, glasses and serving temperature as above, one glass for the 3-digit coded wine and one glass for the pivot wine, marked "P". Panelists were asked to smell and taste the pivot first, then the coded wine. Then they had to check the boxes "more than Pivot" corresponding to the attributes they found to be more pronounced in the coded wine than in the Pivot. The APT questionnaire was made as identical as possible to the CATA questionnaire, see Supplemental Table 5, with consent form on one side, and instructions plus scoring table on the other. We used a similar table presentation and only one box per wine and per attribute, to be checked when the attribute was considered as "more" intense than in the pivot. The unchecked box meant "less than Pivot" or "similar to Pivot". At the end, they were asked to give a hedonic score as described in the CATA paragraph. Between each wine, panelists had to rinse their glass and their mouth with water.

### 2.4. Data treatment

Raw data typing-in was performed by students using a shared spreadsheet table. An example is available online. For a coordinated typing-in session, the filled questionnaires were distributed equally among the students, then each student typed-in the data on a free section of the online spreadsheet, leaving 100 blank lines between two students, to avoid overwriting on each other. These blank lines were later suppressed. An additional instruction was 'not to delete any line' during the typing-in process. Students were typing-in "1" for each box that was checked. "Zeros" were added later on the Excel spreadsheet, once uploaded. The advantage of typing-in server and panelist codes is for later checks for mistyping. Once uploaded in Excel the data was converted to a file like 'DataForCA.csv', using semicolons as separators. Moreover, the names of the two first columns of this data file must be "Product" and "NumPanel" for, respectively, the names of the tested products and the numbers corresponding to the panelists (from 1 to the maximum number of panelists). This is critical for the use of the R script performing the correspondence analysis. All files are available as Supplemental files in FilesForCA_CATA.zip and FilesForCA_APT.zip.

For the R analysis, it is important to have both files in the same directory. Each time another CA analysis is run with a new set of data, it is recommended to create a new directory, this allows to keep the same file names and avoid to modify the R script. It is also necessary to upload the FactoMineR R package (R Core Team, 2021; Lê et al., 2008). Confidence ellipses around wines, with a 95% confidence level, were performed with the *ellipseCA* function of *FactoMineR* package, using the default argument method="multinomial", with limitations in interpretation as discussed in Beaulieu et al. (2022).

To estimate the minimal number of panelists needed to obtain a significant Chi-square test of independence, associated with the contingency table built on attributes and wines, we calculated the evolution of the *p*-value of the Chi-squared test depending on the number of panelists. For this purpose, for a dozen of equally spaced numbers of panelists, from 10 to approximately the total number of panelists, the R script draw at random 300 samples of panelists of the corresponding size out of the batch of 63 or 65 tasters, in APT and CATA, respectively.

The analyses of hedonic scores were performed using ANOVAs and Tukey HSD tests for multiple comparisons from stats and agricolae R packages (R core Team, 2021; De Mendiburu, 2020). Raw data and script for ANOVA analyses are available as Supplemental files (FilesForANOVA.zip).

## 3. RESULTS

### 3.1. CATA analysis

The CA map resulting from the CATA session is shown in Figure 1A. Axis 1, representing 54% of the variability, opposed mainly "fruity", "sweet" and "floral" to other attributes, whereas axis 2, with 25% of the variability, opposed "alcohol" and "strong" to "vegetal" and "light". The CA shows that the *MUSCAT* wine was associated with "floral", "fruity" and "sweet". CHAS-SPUR was associated to "alcohol" and "strong", whereas CHAS-MP was associated with "vegetal" and "light". The identical wine, presented twice under different codes, *VIOGNIER1* and *VIOGNIER2* were described as an identical wine, which they were, very close on the graph with both 95% confidence ellipses overlapping. Finally, *ITALIA* was placed in the center, associated mainly with "spicy" and "linger" for lingering.

The Figure 1B represents, with the CATA method, the evolution of the p-value of the Chi-square test for independence, associated with the contingency table built on attributes and wines, depending on the number of tasters (panelists). This means that when the p-value drops below 0.05 there is a strong link between attributes and wines, and that the analysis is quite robust. It was the case with a panelist number above 50.

The hedonic scores were analyzed by ANOVA and Tukey's post-hoc tests, the detailed results are shown in Supplemental Figure 1. The *MUSCAT* wine was the only significantly preferred wine with an average hedonic score of 6.4, whereas the other scores were 5.5, 5.4, 5.2, 5.2 and 5.1, for *ITALIA*, *VIOGNIER1*, *CHAS-SPUR*, *CHAS-MP* and *VIOGNIER2*, respectively.

### 3.2. APT analysis

The CA map resulting from the APT session is shown in Figure 2A. The axis 1, representing 47% of the variability, opposed mainly "strong", "alcohol" and "spicy" to "light", "fresh", "acid" and "bland", whereas the axis 2, with 25% of the variability, opposed "short" "woody" and "bland" to "fruity" "vegetal" and "floral". Both *VIOGNIER1* and *VIOGNIER2* were analyzed as quite similar, but not as close as in the CATA session, and *MUSCAT* was again associated with "fruity", floral" and "sweet". *CHAS-SPUR* and *CHAS-MP* were again opposed, with *CHAS-SPUR* being associated with "strong", "alcohol" and "spicy", and *CHAS-MP* with "short", "fresh" and "light". With this APT session *ITALIA* was associated with "acid" and "light". But the obvious difference, between the CATA and the APT CA analyzes, is that the 95% confidence ellipses are quite larger in the APT, illustrating a larger variability of the panelist perceptions using the method. Moreover, the p-value of the Chi-square test for independence calculated as a function on the number of panelists never dropped below 0.8 for up to 60 panelists, showing no strong relation between attributes and wines (Figure 2B). The analyses of hedonic scores gave no significant difference (unshown data).

## 4. DISCUSSION

The CATA method associated with CA analyses were efficient to describe white wines with a rather small panel of untrained students, around 50 persons (Figure 1B), confirming that CATA is a very robust method, as illustrated by the large set of studies using this method (Jaeger et al., 2020, and refs herein). The quality of our CATA analysis is proved by several items: **a)** the identical wine, *VIOGNIER*, presented twice under different codes, was analyzed as similar; **b)** the *CHAS-SPUR* (stronger alcohol content and less acidity) and *CHAS-MP* (weaker alcohol percent and stronger acidity), see Supplemental Table 1 for chemical analyses, were opposed logically (Figure 1A); **c)** the *MUSCAT* wine, made of a cultivar with terpenic aromas, typical of some flowers and fruits (Ribéreau-Gayon et al., 1975), was described as "floral" and "fruity" (Figure 1A).

The APT method associated with CA analyses gave weaker results than CATA. We will try to analyze them in comparison with rather positive results we obtained previously with a similar method (Beaulieu et al, 2022). In this former study, the p-value of the Chi-square test dropped below 0.05 for 50 panelists in the case of white wine analyses. The main differences, between the present study and the former one, were: **a)** Here we respected the CATA rule of randomizing the presentation order of the attributes, something we did not perform in Beaulieu's study. Indeed, in this former study, we kept the same order for the attributes in all questionnaires, from olfactive to gustative ones, as the wine professionals do; however, this randomization of attribute order in the questionnaire does not seem to limit the CATA analyses. **b)** In the former study, we presented only four whites, compared to six here, and there were only 13 attributes, compared to 17 here; this could generate slightly higher sensory fatigue. **c)** Finally, in the former study, the questionnaire was harboring the "less" and "more" boxes (than the pivot) for each attribute. Whereas we decided to harbor only one box to tick here, the "more" than the pivot, to make the APT questionnaire visually similar to the CATA one. And the APT data process was similar to CATA, as a checked box was converted to "1" in the typing-in process of the result table, in each method. This led us to analyze the absence of checking this box as "less" or "equal" to the pivot, creating a slight bias. Indeed, the presence of the "less" box creates a forced choice between "more" and "less".

In a recent study by Wang et al. (2023), a similar comparison between CATA and APT was performed, the equivalent of APT was called Pivot-CATA. In their study, they found that the Pivot-CATA decreased the size of the 95% confidence ellipses compared to CATA, contrary to our results, but they also found that the sum of CA axes, which shows the proportion of variance explained by horizontal and vertical dimensions, was greater in CATA than in Pivot-CATA, a similar result to what we found here. In a previous study, Miraballes et al. (2018) found similar results between Pivot-CATA and Pivot© Profile, but they did not perform comparisons with CATA.

Finally, we think having twice more samples to taste in the pivot tests, than in the CATA tests, could create sensory fatigue more rapidly, thus limiting the efficiency of the tests with consumers, as mentioned by Francis and Williamson (2015), particularly with a high number of attributes to check.

## 5. CONCLUSION

The main interest of this set of experiments was to develop a rapid and robust method to teach the students to perform it, using free online tools. We believe that it could also be useful for a large set of professionals using sensory analyses punctually, or teaching sensory methods in extension programs.

CATA proved to be a robust analysis, easy to understand by students organizing the tests rapidly, and allowing good descriptions despite the lack of wine expertise by students who were panelists, when using attributes that they generated in a preliminary session.

Two limitations of the pivot methods seem to be: **a)** the heavier set-up, i.e. in our study, two glasses per panelist, instead of one in the CATA method; **b)** the limitation of the pivot methods to foods or products that can be mixed to equal proportions to create the pivot, this mixing step being easier for liquid products, than for solid products (fruits, chocolate bars…).

**Acknowledgements:** The authors thank all panel students for their time and participation, and the Pech Rouge experimental station for wine samples. Thanks to the Région Occitanie and Toulouse INP for funding part of the study via the Défi Clé Vinid'Occ.

**Author contributions:** Sylvain Nougarede and Alice Diot: organizing and running tasting panels, and data acquisition. Elie Maza: statistical treatment, Alain Samson: wine making, Soline Caillé and Olivier Geffroy: expertise in sensory methods, Valerie Olivier and Christian Chervin: securing the funding, Christian Chervin: organizing tasting panels and writing the original draft. E.M., A.S., O.G. and C.C. performed the conceptualization of the study and article, involving S.N. and A.D. in these steps. All co-authors edited the manuscript.

**Conflict of interest:**

Potential conflict of interest with some French research teams with which we worked previously.

**Data availability statement:**

FilesForCA.zip and FilesForANOVA.zip contain raw data and R scripts, and they are available on demand (christian.chervin (a) ensat.fr).

**ORCID**

Elie Maza 0000-0002-7351-6345

Valérie Olivier-Salvagnac 0000-0002-4724-2619

Olivier Geffroy 0000-0002-8655-5669

Christian Chervin 0000-0002-0913-6815

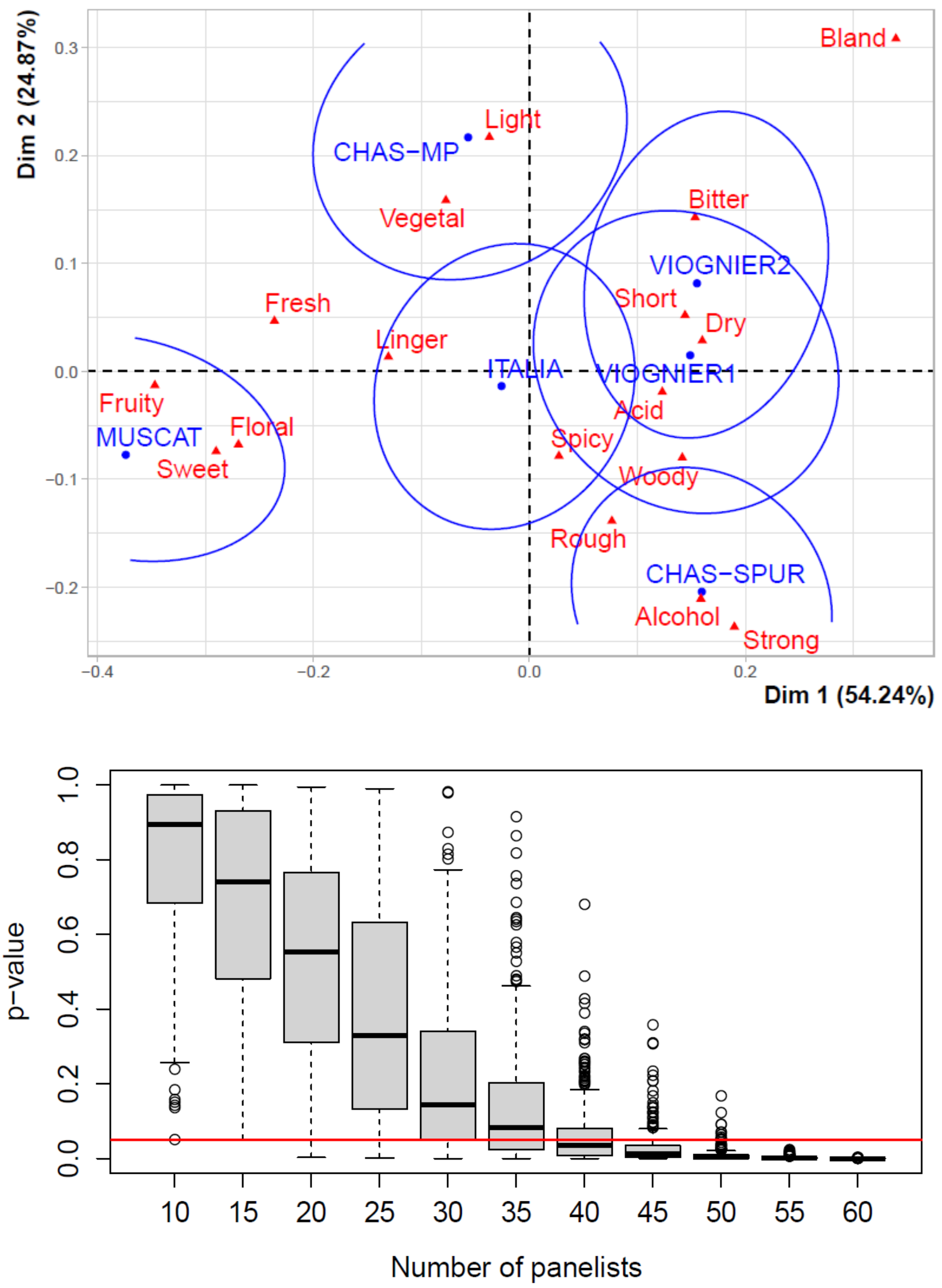

A]

B]

**<u>Figure 1</u>**: **A]** Correspondence Analyse after a **CATA session** with white wines, with 65 panelists; wine labels in blue, attribute labels in red; the ellipses represent the 95% interval. **B]** p-value of the chi-squared test for independence of attributes and wines, as a function of the number of panelists, the red line corresponds to $p = 0.05$.

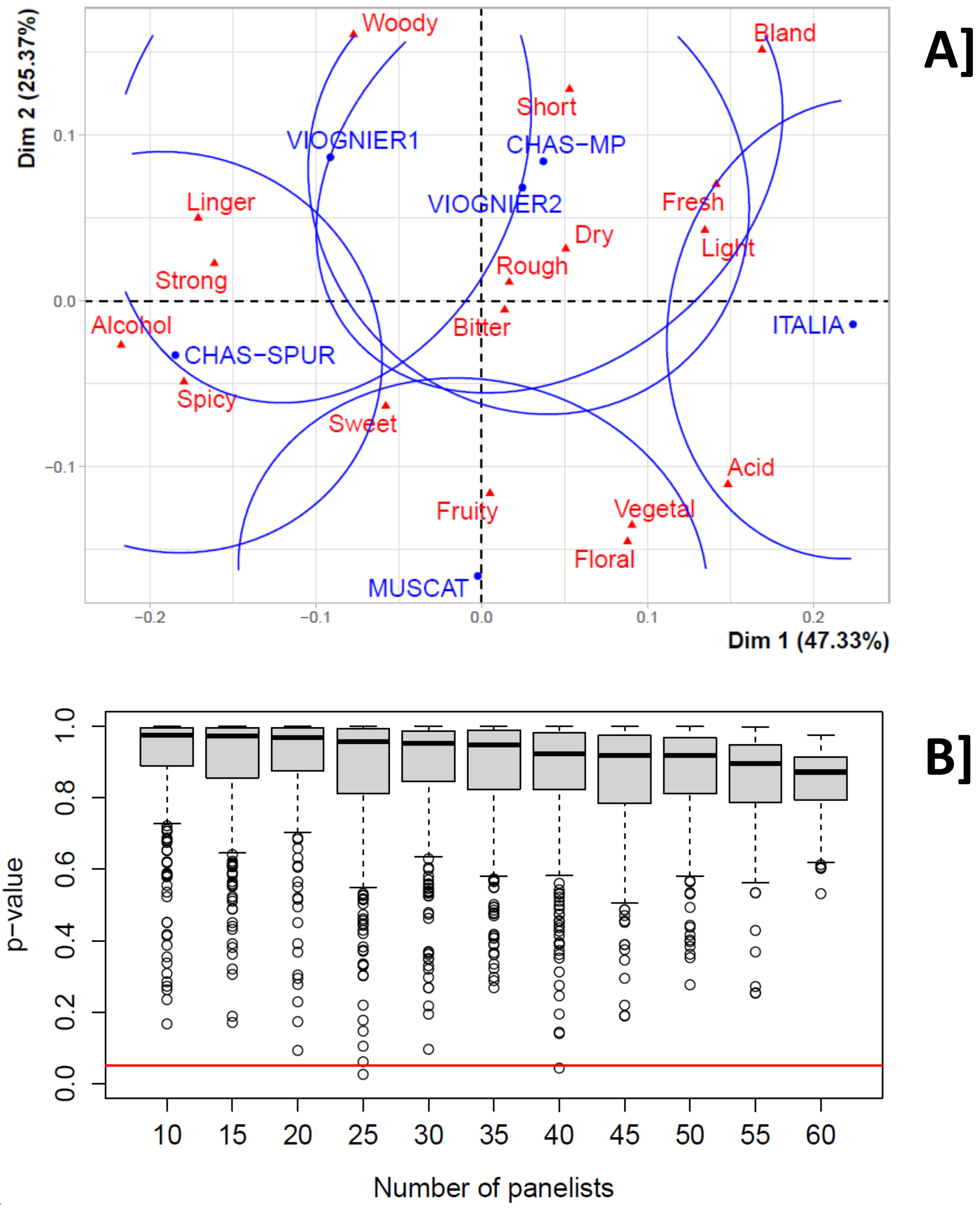

**<u>Figure 2</u>**: **A]** Correspondence Analyse after a **APT session** with white wines, with 63 panelists; wine labels in blue, attribute labels in red; the ellipses represent the 95% interval. **B]** p-value of the chi-squared test for independence of attributes and wines, as a function of the number of panelists, the red line corresponds to $p = 0.05$.

**Supplemental Table 1:** Wine characteristics and cultivars

| Wine | Alcohol % (v/v) | Volatile acidity g/L $H_2SO_4$ | pH | Titratable acidity g/L $H_2SO_4$ | free $SO_2$ mg/L | total $SO_2$ mg/L | OD 280 nm | OD 420 nm | L* | a* | b* | Chroma | Hue angle | year | cultivar | last cross | pruning |
|---|---|---|---|---|---|---|---|---|---|---|---|---|---|---|---|---|---|
| CHAS-SPUR | 14.94 | 0.20 | 3.33 | 3.56 | 14 | 63 | nd | 0.088 | 98.9 | -0.764 | 6.519 | 6.56 | 96.68 | 2021 | 3159-2-12 | Chasan | spur |
| CHAS-MP | 12.92 | 0.11 | 3.36 | 3.23 | 22 | 76 | nd | 0.086 | 98.9 | -0.822 | 6.127 | 6.18 | 97.64 | 2021 | 3159-2-12 | Chasan | minimal |
| ITALIA | 11.82 | 0.20 | 3.23 | 4.61 | 20 | 87 | nd | 0.051 | 99.5 | -0.811 | 4.258 | 4.33 | 100.78 | 2021 | 3196-57 | Italia | spur |
| MUSCAT | 13.65 | 0.27 | 3.01 | 4.07 | 22 | 101 | 5.91 | 0.062 | 99.2 | -0.476 | 4.801 | 4.82 | 95.66 | 2021 | 3197-81 | Muscat Hambourg | spur |
| VIOGNIER | 12.04 | 0.40 | 3.50 | 4.64 | 16 | 58 | 6.66 | 0.113 | 98.6 | -1.007 | 8.394 | 8.45 | 96.84 | 2021 | Viognier | - | spur |

Analyses performed as in Beaulieu et al. (2022)
"nd" stands for "not determined"

**Supp. Table 2:** Panelist details

| | 'Attribute' session | | 'CATA' session | | 'APT' session | |
|---|---|---|---|---|---|---|
| ***Age*** | ***Average*** | ***Std error*** | ***Average*** | ***Std error*** | ***Average*** | ***Std error*** |
| **Year old** | 21.7 | 1.2 | 21.2 | 1.4 | 21.6 | 1.5 |
| ***Gender*** | ***n*** | ***%*** | ***n*** | ***%*** | ***n*** | ***%*** |
| **Female** | 18 | 60 | 40 | 62 | 42 | 67 |
| **Male** | 12 | 40 | 25 | 38 | 21 | 33 |

The total panel was composed of 67 students: 42 women and 25 men, with an average age of 21.6 years.
86.5% of the total panel participated to both CATA and APT panels.

**Supplemental Table 3:** Occurrence of the attributes after lemmatisation and triangulation, then translation (the initial set of attributes was in French)

| Attributes | % of occurrence | | List of the 17 attributes kept for CATA and APT comparison | | Random orders to present the attributes in questionnaires | | | | | | | | | |
|---|---|---|---|---|---|---|---|---|---|---|---|---|---|---|
| Fruity | 16.54 | 1 | Acid | see Supp Table 6 for random drawing of | **Order 1** | **Order 2** | **Order 3** | **Order 4** | **Order 5** | **Order 6** | **Order 7** | **Order 8** | **Order 9** | **Order 10** |
| Sweet | 14.13 | 2 | Alcohol | the attributes | Bitter | Fresh | Fruity | Short | Bitter | Bland | Spicy | Acid | Rough | Linger |
| Acid | 11.7 | 3 | Bitter | | Dry | Linger | Linger | Fruity | Vegetal | Floral | Light | Fruity | Fresh | Bitter |
| Dry | 8.76 | 4 | Bland | | Spicy | Fruity | Dry | Vegetal | Rough | Vegetal | Rough | Dry | Alcohol | Rough |
| Alcohol | 7.51 | 5 | Dry | | Fresh | Light | Short | Rough | Strong | Spicy | Bland | Rough | Woody | Floral |
| Light | 6.52 | 6 | Floral | | Short | Vegetal | Alcohol | Bitter | Floral | Bitter | Short | Floral | Bitter | Fruity |
| Bitter | 5.86 | 7 | Fresh | | Sweet | Spicy | Light | Linger | Alcohol | Strong | Bitter | Light | Bland | Fresh |
| Spicy | 5.64 | 8 | Fruity | | Linger | Sweet | Woody | Bland | Spicy | Fruity | Floral | Linger | Floral | Alcohol |
| Bland | 3.91 | 9 | Light | | Fruity | Strong | Fresh | Fresh | Dry | Acid | Vegetal | Sweet | Vegetal | Acid |
| Rough | 3.33 | 10 | Linger | | Light | Short | Vegetal | Dry | Short | Fresh | Woody | Bland | Strong | Strong |
| Strong | 2.65 | 11 | Rough | | Woody | Floral | Sweet | Light | Fruity | Linger | Dry | Strong | Light | Sweet |
| Short | 2.44 | 12 | Short | | Floral | Alcohol | Bland | Acid | Sweet | Light | Alcohol | Short | Linger | Dry |
| Fresh | 1.99 | 13 | Spicy | | Acid | Acid | Rough | Floral | Acid | Rough | Linger | Fresh | Fruity | Vegetal |
| Floral | 1.54 | 14 | Strong | | Bland | Rough | Bitter | Alcohol | Light | Dry | Strong | Woody | Short | Spicy |
| Vegetal | 1.1 | 15 | Sweet | | Strong | Woody | Acid | Strong | Linger | Short | Fresh | Alcohol | Spicy | Light |
| Linger | 1.1 | 16 | Vegetal | | Vegetal | Bland | Spicy | Spicy | Bland | Woody | Sweet | Spicy | Sweet | Woody |
| Woody | 1.1 | 17 | Woody | | Alcohol | Bitter | Strong | Woody | Woody | Alcohol | Fruity | Vegetal | Dry | Short |
| Iodine | 0.88 | | | | Rough | Dry | Floral | Sweet | Fresh | Sweet | Acid | Bitter | Acid | Bland |
| Undergrowth | 0.88 | | | | | | | | | | | | | |
| Neutral | 0.66 | | | | | | | | | | | | | |
| Mineral | 0.66 | | | | | | | | | | | | | |
| Smelly | 0.66 | | | | | | | | | | | | | |
| Thick | 0.44 | | | | | | | | | | | | | |

**Supplemental Table 4:** Example of CATA questionnaire

(o3)

# CATA questionnaire

**Surname :** ............................ **Age :** ....................

**Name :** ........................... ***Panelist code*** *(4 first let. of surname + 4 first let. of name):* ..........................

***Server code*** *(given by the server):* .........................

***Instructions:*** *You will taste six wines, one by one. Check all the boxes that correspond to the attributes of the coded wine. On the last line, give a score between 1 and 9 for your overall appreciation of the coded wine (1 = I don't like it and 9 = I like it very much).*

| **Attributes** | **Wine code: ......** | **Wine code: ......** | **Wine code: ......** | **Wine code: ......** | **Wine code: ......** | **Wine code: ......** | **Attributes** |
|---|---|---|---|---|---|---|---|
| Fruity | ☐ | ☐ | ☐ | ☐ | ☐ | ☐ | Fruity |
| Linger | ☐ | ☐ | ☐ | ☐ | ☐ | ☐ | Linger |
| Dry | ☐ | ☐ | ☐ | ☐ | ☐ | ☐ | Dry |
| Short | ☐ | ☐ | ☐ | ☐ | ☐ | ☐ | Short |
| Alcohol | ☐ | ☐ | ☐ | ☐ | ☐ | ☐ | Alcohol |
| Light | ☐ | ☐ | ☐ | ☐ | ☐ | ☐ | Light |
| Woody | ☐ | ☐ | ☐ | ☐ | ☐ | ☐ | Woody |
| Fresh | ☐ | ☐ | ☐ | ☐ | ☐ | ☐ | Fresh |
| Vegetal | ☐ | ☐ | ☐ | ☐ | ☐ | ☐ | Vegetal |
| Sweet | ☐ | ☐ | ☐ | ☐ | ☐ | ☐ | Sweet |
| Bland | ☐ | ☐ | ☐ | ☐ | ☐ | ☐ | Bland |
| Rough | ☐ | ☐ | ☐ | ☐ | ☐ | ☐ | Rough |
| Bitter | ☐ | ☐ | ☐ | ☐ | ☐ | ☐ | Bitter |
| Acid | ☐ | ☐ | ☐ | ☐ | ☐ | ☐ | Acid |
| Spicy | ☐ | ☐ | ☐ | ☐ | ☐ | ☐ | Spicy |
| Strong | ☐ | ☐ | ☐ | ☐ | ☐ | ☐ | Strong |
| Floral | ☐ | ☐ | ☐ | ☐ | ☐ | ☐ | Floral |
| **Score 1 to 9** | | | | | | | **Score 1 to 9** |

**Supplemental Table 5:** Example of APT questionnaire

(o3)

# APT questionnaire

**Surname :** ………………………  **Age :** ……………….

**Name :** ………………………  ***Panelist code*** *(4 first let. of surname + 4 first let. of name): ………………………*

***Server code*** *(given by the server): ………………………*

***Instructions:*** *1) You will taste 6 coded wines in comparison with a pivot wine coded "P". Taste the Pivot first, then the coded wine. For each attribute, you will check the "PLUS" box or nothing. Example: if you perceive the "456" wine as more fruity than the "P" wine, you check the box. If you perceive the "456" wine to be less fruity (or as fruity) as the "P" wine, you leave the box unchecked. On the last line, give a score between 1 and 9 for your overall appreciation of the coded wine (1 = I don't like it and 9 = I like it very much).*

| | **Wine code: ……** | **Wine code: ……** | **Wine code: ……** | **Wine code: ……** | **Wine code: ……** | **Wine code: ……** | |
|---|---|---|---|---|---|---|---|
| **Attributes** | more than "P" | more than "P" | more than "P" | more than "P" | more than "P" | more than "P" | **Attributes** |
| Fruity | ☐ | ☐ | ☐ | ☐ | ☐ | ☐ | Fruity |
| Linger | ☐ | ☐ | ☐ | ☐ | ☐ | ☐ | Linger |
| Dry | ☐ | ☐ | ☐ | ☐ | ☐ | ☐ | Dry |
| Short | ☐ | ☐ | ☐ | ☐ | ☐ | ☐ | Short |
| Alcohol | ☐ | ☐ | ☐ | ☐ | ☐ | ☐ | Alcohol |
| Light | ☐ | ☐ | ☐ | ☐ | ☐ | ☐ | Light |
| Woody | ☐ | ☐ | ☐ | ☐ | ☐ | ☐ | Woody |
| Fresh | ☐ | ☐ | ☐ | ☐ | ☐ | ☐ | Fresh |
| Vegetal | ☐ | ☐ | ☐ | ☐ | ☐ | ☐ | Vegetal |
| Sweet | ☐ | ☐ | ☐ | ☐ | ☐ | ☐ | Sweet |
| Bland | ☐ | ☐ | ☐ | ☐ | ☐ | ☐ | Bland |
| Rough | ☐ | ☐ | ☐ | ☐ | ☐ | ☐ | Rough |
| Bitter | ☐ | ☐ | ☐ | ☐ | ☐ | ☐ | Bitter |
| Acid | ☐ | ☐ | ☐ | ☐ | ☐ | ☐ | Acid |
| Spicy | ☐ | ☐ | ☐ | ☐ | ☐ | ☐ | Spicy |
| Strong | ☐ | ☐ | ☐ | ☐ | ☐ | ☐ | Strong |
| Floral | ☐ | ☐ | ☐ | ☐ | ☐ | ☐ | Floral |
| **Score 1 to 9** | | | | | | | **Score 1 to 9** |

**Supplemental Table 6** : Random draw of attributes

**List of the 17 attributes kept for CATA and APT comparison**

Excel files available on demand (christian.chervin (a) ensat.fr)

| | | | | |
|---|---|---|---|---|
| 1 | Acid | 0.47457544 | 9 | Dry |
| 2 | Alcohol | 0.90007122 | 17 | Strong |
| 3 | Bitter | 0.27189569 | 6 | Vegetal |
| 4 | Bland | 0.61992847 | 12 | Woody |
| 5 | Dry | 0.08678929 | 1 | Floral |
| 6 | Floral | 0.26826941 | 5 | Bitter |
| 7 | Fresh | 0.69934062 | 13 | Spicy |
| 8 | Fruity | 0.85313891 | 16 | Short |
| 9 | Light | 0.73209592 | 14 | Acid |
| 10 | Linger | 0.74709298 | 15 | Rough |
| 11 | Rough | 0.49527028 | 10 | Sweet |
| 12 | Short | 0.45275129 | 8 | Bland |
| 13 | Spicy | 0.40904023 | 7 | Fresh |
| 14 | Strong | 0.14017616 | 2 | Light |
| 15 | Sweet | 0.54301846 | 11 | Linger |
| 16 | Vegetal | 0.16014265 | 3 | Fruity |
| 17 | Woody | 0.22906006 | 4 | Alcohol |

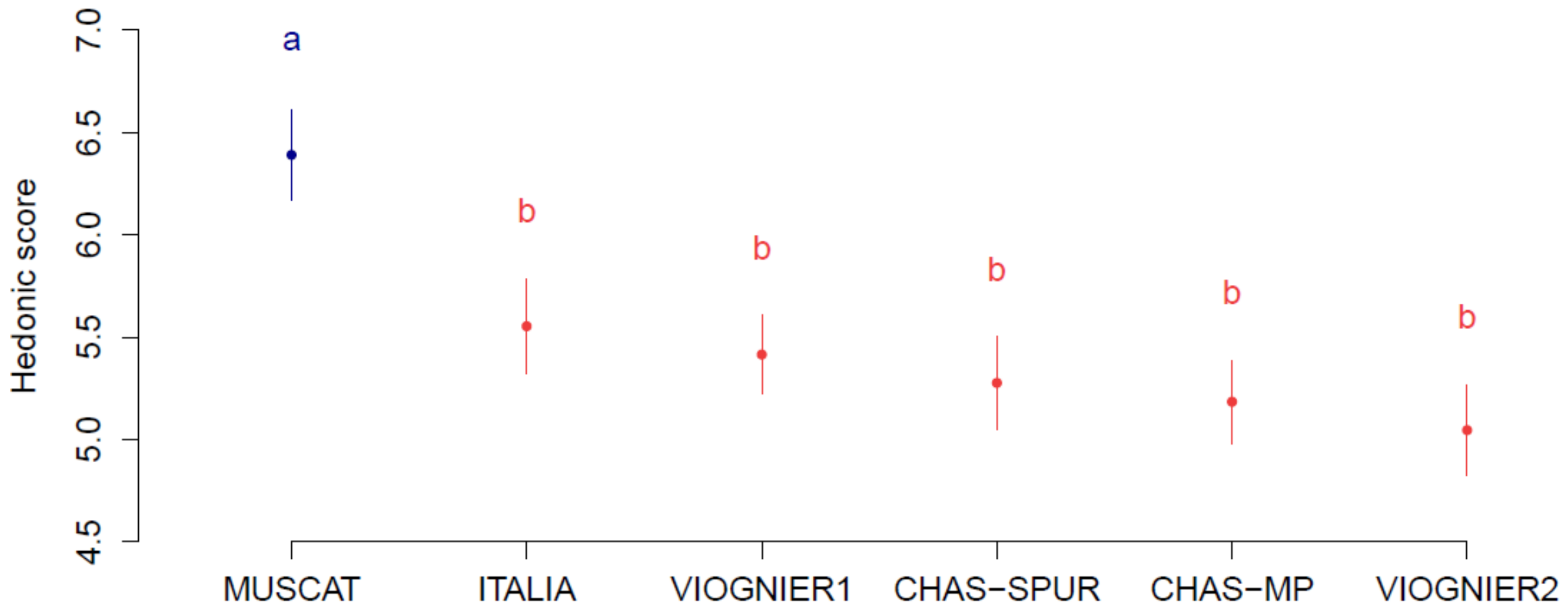

**Supplemental Figure 1**: Mean hedonic scores given to white wines by 65 panelists at the end of a CATA session, error bars show SE, different letters highlight significant differences by Tukey's HSD ($p < 0.05$).